\documentclass[aps,pra,groupedaddress,showpacs]{revtex4}

\begin{document}

\draft
\title{Quantum entanglement as an interpretation of bosonic character
in composite two-particle systems}

\author{C. K. Law}

\affiliation{Department of Physics, The Chinese University of Hong
Kong, Shatin, Hong Kong SAR, China}


\date{\today}


\begin{abstract}
We consider a composite particle formed by two fermions or two
bosons. We discover that composite behavior is deeply related to
the quantum entanglement between the constituent particles. By
analyzing the properties of creation and annihilation operators,
we show that bosonic character emerges if the constituent
particles become strongly entangled. Such a connection is
demonstrated explicitly in a class of two-particle wave functions.
\end{abstract}


\maketitle

In Nature many particles are composite particles formed by two or
more fundamental fermions. Atoms, molecules and nuclei are
familiar examples of composite particles. As long as physical
processes do not reveal the internal structure, a composite
particle as a whole can be treated approximately as a point-like
boson (fermion) if the number of constituent fermions is even
(odd). Intuitively, the more compact the composite particle is,
the better it behaves as a pure boson or fermion. This raises
interesting questions how the degree of compositeness is
quantified, and what physical effects would appear if the internal
structure becomes apparent. In the case of particles formed by two
fermions, several authors have addressed the compositeness effects
of atoms in Bose-Einstein condensates \cite{Avancini,Rombous} and
electron-hole pairs in semiconductor \cite{Rombous,Combescot}. In
particular, Combescot and Tanguy have indicated the deviation of
purely bosonic behavior, based on the properties of the creation
operator associated with composite particles \cite{Combescot}.

The issue of compositeness also appears in particles made of
bosons. An important example of this kind is biphoton states
generated by optical spontaneous parametric down conversion. Such
highly correlated photons have been employed in fundamental
experiments, e.g., to test the violation of Bell's inequality
\cite{bell1}, and to realize Einstein-Podolsky-Rosen paradox
\cite{howell}. Recent experiments have also explored the `binding'
effects revealed from the de Broglie wavelength of biphotons
\cite{dbwave,Edamatsu}. Interestingly, although the constituent
photons are non-interacting and they can be spatially separated,
the composite nature of biphotons can still be detected by
suitable interference schemes. This suggests that the concept of
compositeness is not limited to mechanically bounded particles in
position space.

In this paper we examine the composite representation of
two-particle systems from the view point of quantum information.
We show that quantum entanglement provides an understanding of the
origin of composite behavior. As we shall see below, the degree of
entanglement between constituent particles determines how close a
composite particle behave as a pure boson. This implies an
interesting picture that constituent particles are somehow bounded
by quantum entanglement. Mechanical binding forces are not
essential, they serve only as physical means to enforce quantum
correlations. Since the constituent particles can be correlated in
many different ways, the composite representation is not limited
to position space or momentum space. In this paper we will
indicate the fundamental role of entanglement, based on the
properties of creation and annihilation operators associated with
composite particles. In contrast to previous studies that concern
only fermionic constituent particles, we provide a general
analysis that applies to bosonic constituents as well.

To begin with, we consider a composite particle $C$ formed by two
distinguishable particles $A$ and $B$.  Both $A$ and $B$ are
either fermions or bosons. Let $\Psi ({ x}_A ,{ x}_B )$ be the
wave function of the two-particle system, we can always express
the wave function in the Schmidt decomposition form:
\begin{equation}
\Psi ({ x}_A ,{ x}_B ) = \sum\limits_{n=0}^{\infty} {\sqrt
{\lambda _n } } \phi _n^{(A)} ({ x}_A )\phi _n^{(B)} ({ x}_B )
\end{equation}
where the Schmidt modes $\phi _n^{(A)}$ ($\phi _n^{(B)}$) form a
complete and orthonormal set for particle $A$ ($B$). Specifically,
$\phi _n^{(A)}$ and $\lambda_n$ are defined by the eigenvectors
and eigenvalues of the reduced density matrix of particle $A$, and
$\phi _n^{(B)}$ can be obtained similarly from the reduced density
matrix of particle $B$. Eq. (1) reveals the quantum correlation by
showing the pairing structure explicitly. If the particle $A$
appears in the mode $\phi _n^{(A)}$, then with certainty the
particle $B$ must be in the mode $\phi _n^{(B)}$. The distribution
of $\lambda_n$ provides a measure of entanglement. This is usually
discussed in terms of the entanglement entropy $ E = -
\sum\nolimits_n {\lambda _n} \log _2 \lambda _n $. However, a more
transparent measure of entanglement is to count the `average'
number of Schmidt modes actively involved. The Schmidt number
${\cal K}$ provides this information \cite{reiner,law}:
\begin{equation}
{\cal K} \equiv 1/\sum\limits_{n=0}^{\infty} {\lambda _n^2 }.
\end{equation}
The larger the value of ${\cal K}$ is, the higher the
entanglement. In fact, ${\cal K}$ equals the inverse of the purity
of single-particle density matrix, and it also equals the linear
entropy apart from a constant. A disentangled (product) state
corresponds to ${\cal K}=1$, i.e., only one term in the Schmidt
decomposition. If a Schmidt decomposition contains $M$ terms of
equal weight (i.e., $\lambda_n=1/M$  for $n=1,2,..,M$), then we
have ${\cal K}=M$ which is exactly the number of mode pairs
involved.

In the second quantized representation, Eq. (1) corresponds to a
state generated by a creation operator $c^{\dag}$ acting on the
vacuum. Such a creation operator is defined by,
\begin{equation}
c^{\dag}  = \sum\limits_{n=0}^{\infty} {\sqrt {\lambda _n }
a_n^{\dag} } b _n^{\dag}
\end{equation}
where $a_n^{\dag}$ ($b_n^{\dag}$) is the creation operator of the
particle $A$ ($B$) in the Schmidt mode $\phi _n^{(A)}$ ($
\phi_n^{(B)}$). The commutation relation between $c$ and
$c^{\dag}$ is given by,
\begin{equation}
[ {c,c^{\dag} } ] = 1 + s \Delta,
\end{equation}
where $s=+1$ if $A$ and $B$ are bosons, $s=-1$ if $A$ and $B$ are
fermions. The operator $\Delta$ in Eq. (4) is defined by,
\begin{equation}
\Delta= \sum\limits_{n=0}^{\infty} {\lambda _n } \left( {a
_n^{\dag} a _n + b _n^{\dag} b _n } \right)
\end{equation}
which has non-zero matrix elements, depending on the states
involved. Therefore $c$ and $c^{\dag}$ are not strictly bosonic
operators.

To examine the properties of $c$ and $c^{\dag}$, we consider a
system containing two or more $C$ particles. Let us define the
$N-$particle states by,
\begin{equation}
\left| N \right\rangle  \equiv \chi _N^{-1/2} \frac{{c^{\dag N}
}}{{\sqrt {N!} }}\left| 0 \right\rangle
\end{equation}
where $\chi_N$ is a normalization constant such that $ \langle {N}
| {N} \rangle  = 1$. In order to test how good the operator $c$
behaves as a bosonic annihilation operator, we need to determine
how the operator acts on $\left| N \right\rangle$. This is given
by a general equation:
\begin{equation}
c\left| N \right\rangle  = \alpha _N \sqrt N \left| {N - 1}
\right\rangle  + \left| {\varepsilon _N } \right\rangle,
\end{equation}
where $\alpha _N$ is a constant, and the correction term $\left|
{\varepsilon _N } \right\rangle$ is orthogonal to $ \left| {N - 1}
\right\rangle$. Such a correction term is necessary because the
set of $\left| N \right\rangle$ states (with all $N$) is only a
subset of the entire Hilbert space associated with the constituent
particles. A non-ideal bosonic operator would inevitably cause
transitions into the states not described by (6).

From Eq. (7), the annihilation operator $c$ is bosonic if the
following two conditions are satisfied:
\begin{eqnarray}
 \alpha_N  \to 1 \\
\langle \varepsilon _N | \varepsilon _N \rangle \to 0.
\end{eqnarray}
After some calculations \cite{delta}, for $N>1$, we have
\begin{eqnarray} && \alpha _N  = \sqrt
{\frac{{\chi _N }}{{\chi _{N - 1} }}}
\\
&& \langle \varepsilon _N | \varepsilon _N \rangle  =
 1 - N\frac{{\chi _N }}{{\chi _{N - 1} }}
 + (N - 1)\frac{{\chi _{N + 1} }}{{\chi _N }}.
\end{eqnarray}
Therefore the conditions (8) and (9) are controlled by the ratio
of normalization constants. An ideal composite boson emerges in
the limit $\chi_{N \pm 1}/ \chi_{N} \to 1$.

For convenience, let us write
\begin{equation}
\chi _N  = \left\{ {\begin{array}{*{20}c}
   {\chi _N^F \ \ \ \ \ \ A,B \ \ {\rm are \ \ fermions}}  \\
   {\chi _N^B \ \ \ \ \ \ \ A,B \ \ {\rm are \ \ bosons.}}  \\
\end{array}} \right.
\end{equation}
By carefully counting the states allowed for ferimons and bosons,
it can be shown that $\chi _N^B$ and $\chi _N^F$ are given by
\cite{chi},
\begin{eqnarray}
\chi _N^B = N!  \sum\limits_{p_N \ge p_{N-1} \ge ....\ge p_2 \ge
p_1 } {\lambda _{p_1 } }
 \lambda _{p_2 } ....\lambda _{p_N } \\ \chi _N^F  = N!
\sum\limits_{p_N > p_{N-1} > ....> p_2 > p_1 } {\lambda _{p_1 } }
 \lambda _{p_2 } ....\lambda _{p_N }.
\end{eqnarray}
The calculation of these summations can be quite complicated in
general. In the case particles $A$ and $B$ are fermions,
$\chi_{N}^F$ can be analyzed by methods discussed in Ref. \cite{
Combescot}.  However, exact analytical expressions of $\chi_{N}$
for both fermions and bosons in closed forms are rare.

In this paper we consider a realistic class of two-particle wave
functions that allows exact closed form expressions of $\chi_{N}$.
Such a class of two-particle wave functions is specified by the
Schmidt eigenvalues,
\begin{equation}
\lambda _n  = (1 - z)z^n \ \ \ \ \ \ \ n=0,1,2...
\end{equation}
where the parameter $z$ is defined in the range $0 < z < 1$. This
parameter determines how rapid $\lambda_n$ decreases with $n$. A
representative example of this class of wave functions is (double)
gaussian given by:
\begin{equation}
\Psi (x_A ,x_B ) = {\cal N}e^{ - (x_A  + x_B )^2 /\sigma _c^2 }
e^{(x_A - x_B )^2 /\sigma _r^2 }.
\end{equation}
Here ${\cal N}$ is a normalization constant, and $\sigma_c$ and
$\sigma_r$ are widths along the $x_A+x_B$ and $x_A -x_B$
directions respectively. The Schmidt decomposition of Eq. (16)
gives $\lambda _n$'s in the form of Eq. (15) with
\begin{equation}
z = \left( {\frac{{\sigma _r - \sigma _c }}{{\sigma _r  + \sigma
_c }}} \right)^2.
\end{equation}
The corresponding Schmidt modes in this example are simply the
eigenfunctions of a harmonic oscillator.

To evaluate $\chi_N^B$ and $\chi_N^F$, we let: $p_1 = q_N$, $p_2 =
q_N+q_{N-1}$, $p_3 =q_N+q_{N-1}+q_{N-2}$,....
$p_N=q_N+q_{N-1}+...+q_1$, then Eq. (13) and (14) become,
\begin{eqnarray}
\chi _N^B = N! (1-z)^N  \sum\limits_{q_1  = 0}^\infty
\sum\limits_{q_{2} = 0}^{\infty } {.....} \sum\limits_{q_{N-1}  =
0}^{\infty } \sum\limits_{q_N  = 0}^{\infty }
z^{q_1+2q_2+3q_3+...+Nq_N}
\\
\chi _N^F = N! (1-z)^N  \sum\limits_{q_1  = 1}^\infty
\sum\limits_{q_{2} = 1}^{\infty } {.....} \sum\limits_{q_{N-1}  =
1}^{\infty } \sum\limits_{q_N  = 0}^{\infty }
z^{q_1+2q_2+3q_3+...+Nq_N}.
\end{eqnarray}
These summations can now be carried out easily,
\begin{eqnarray}
&& \chi _N^B  = \frac{{N!(1 - z)^N }}{{(1 - z)(1 - z^2 )....(1 -
z^N )}}
\\
&& \chi _N^F  = \frac{{N!z^{N(N - 1)/2} (1 - z)^N }}{{(1 - z)(1 -
z^2 )....(1 - z^N )}}.
\end{eqnarray}
Hence the normalization ratios are given by,
\begin{eqnarray}
&& \frac{{\chi _{N + 1}^B }}{{\chi _N^B }} = \frac{{(N+1)(1 -
z)}}{{(1 - z^{N + 1} )}}
\\
&& \frac{{\chi _{N + 1}^F }}{{\chi _N^F }} = z^{N} \frac{{(N+1)(1
- z) }}{{(1 - z^{N + 1} )}}.
\end{eqnarray}
From Eq. (7) and (10), the normalization ratio determines the
modification of Bose enhancement factor. The results Eq. (22) and
(23) indicate that $\chi_{N+1}^B/\chi_{N}^B > 1$ and
$\chi_{N+1}^F/\chi_{N}^F < 1$. The difference between the two
types of constituents can be understood, because bosons tend to
stay together in the same state, while fermions do the opposite
due to the exclusion principle. However, quantum statistics of
constituent particles becomes less important as $z$ approaches
one, where the composite particle become a pure boson. In the
example of double gaussian wave function (16), the $z \to 1$ limit
correspond to the cases $\sigma_r \gg \sigma_c$ or $\sigma_c \gg
\sigma_r$. We remark that that the particle number $N$ is a key
factor in Eq. (22) and (23). A larger number of particle number
would require the value of $z$ to be closer to one in order to
maintain the normalization ratio.

Now we can make an explicit connection with quantum entanglement.
For the Schmidt eigenvalues given by (15), the Schmidt number
$\cal K$ defined in Eq. (2) takes the form:
\begin{equation}
{\cal K} = \frac{{1 - z^2 }}{{(1 - z)^2 }}
\end{equation}
which is a monotonic increasing function in the range $0<z<1$. By
expressing $z$ in terms of ${\cal K}$, both
$\chi_{N+1}^B/\chi_{N}^B$ and $\chi_{N+1}^F/\chi_{N}^F$ can be
directly related to the degree of quantum entanglement. As ${\cal
K}$ increases, we find that both $\chi_{N+1}^B/\chi_{N}^B$ and
$\chi_{N+1}^F/\chi_{N}^F$ approach one monotonically for the
entire range of ${\cal K}$. In particular, it can be shown that
for ${\cal K} \gg N$,
\begin{equation}
\chi_{N+1}/\chi_{N} \approx 1 + s N/{\cal K}
\end{equation}
where $s$ defined in Eq. (4). Since the value of ${\cal K}$
corresponds to an effective number of Schmidt modes \cite{reiner},
bosonic particle description is valid when {\em the effective
number of Schmidt modes is much greater than the total number of
composite particles}.

The discussion above indicates that composite behavior strictly
depends on the degree of quantum entanglement, at least for the
class of two-particle wave functions with Schmidt eigenvalues
specified by Eq. (15). We note that Eq. (15) corresponds to a wide
variety of wave functions, because Schmidt modes can be chosen
from any discrete, complete and orthogonal set of basis functions.
Indeed, Schmidt modes do not play any role in determining
$\chi_N$. It is the distribution of $\lambda _n$ that controls the
compositeness of the particle. As long as the Schmidt eigenvalues
are specified by (15), different forms of Schmidt modes correspond
to the same degree of compositeness. This also shares the same
feature with quantum entanglement.

To summarize, our work here examine the foundation of composite
representation of two-particle systems. Within the class of
two-particle wave functions, we show that the origin of composite
behavior is ultimately related to quantum entanglement between
constituent particles. Therefore composite representation can be
applied to strongly entangled particles, which are not limited to
mechanically bounded systems. Finally, we remark that the wave
functions satisfying Eq. (15) are not completely general, but
owing to flexibility of the parameter $z$, these functions may be
used to approximate wave functions in actual situations. This
suggests that the relation between composite behavior and the
degree of quantum entanglement can be quite general. However, a
full analysis would require the study of Eq. (13) and (14) for
arbitrary distribution of Schmidt eigenvalues, which is a topic
open for future investigations.


\begin{acknowledgments}
The author thanks Prof. M.-C. Chu for discussions. This work is
supported in part by the Research Grants Council of the Hong Kong
Special Administrative Region, China (Project No. 400504 and
Project No. 423701).
\end{acknowledgments}


\begin{references}

\bibitem{Avancini} S. S. Avancini, J. R. Marinelli, and G. Krein,
J. Phys. A. {\bf 36}, 9045 (2003).

\bibitem{Rombous} S. Rombouts, D. Van Neck, K. Peirs, and L.
Pollet, Mod. Phys. Lett. A {\bf 17}, 1899 (2002); S. Rombouts, D.
Van Neck, K. Peirs, and L. Pollet, Europhys. Lett. {\bf 63}, 785
(2003).

\bibitem{Combescot} M. Combescot and C. Tanguy, Europhys. Lett.
{\bf 55}, 390 (2001); M. Combescot, X. Leyronas, and C. Tanguy,
Eur. Phys. J. B. {\bf 31}, 17 (2003); M. Combescot and C. Tanguy,
Europhys. Lett. {\bf 63}, 787 (2003).

\bibitem{bell1} Z. Y. Ou and L. Mandel, Phys. Rev. Lett. {\bf 61},
50 (1988); Y. H. Shih and C. O. Alley, Phys. Rev. Lett. {\bf 61},
2921 (1988).

\bibitem{howell}J. C. Howell, R. S. Bennink, S. J. Bentley,
and R. W. Boyd, Phys. Rev. Lett. {\bf 92}, 210403 (2004).

\bibitem{dbwave} J. Jacobson, G. Bjork, I. Chuang,
and Y. Yamamoto, Phys. Rev. Lett. {\bf 74}, 4835 (1995).

\bibitem{Edamatsu} E. J. S. Fonseca, C. H. Monken, and S. Padua,
Phys. Rev. Lett. {\bf 82}, 2868 (1999); E. J. S. Fonseca, Z.
Paulini, P. Nussenzveig, C. H. Monken, and S. Padua, Phys. Rev. A
{\bf 63}, 043819 (2001); K. Edamatsu, R. Shimizu, and T. Itoh,
Phys. Rev. Lett. {\bf 89}, 213601 (2002).

\bibitem{reiner}  R. Grobe, K. Rz\c{a}\.zewski and J.H. Eberly,
J. Phys. B {\bf 27}, L503 (1994).

\bibitem{law} C. K. Law and J. H. Eberly, Phys. Rev. Lett.
{\bf 92}, 127903 (2004).

\bibitem{delta} The derivation of Eq. (11) requires the
expectation value $\langle N | \Delta |N \rangle$, which can be
obtained by using the algebraic method in Ref. [3].

\bibitem{chi} In deriving Eq. (13), we write
\begin{eqnarray}
c^{\dag N} \left| 0 \right\rangle  = \sum\limits_{k_N  \ge k_{N -
1} \ge ..... \ge k_1 }^{} {\sqrt {\lambda _{k_1 } \lambda _{k_2 }
......\lambda _{k_N } } } F\left( {k_1,k_2,...,k_N } \right)\left|
{k_1,k_2,...,k_N } \right\rangle \nonumber
\end{eqnarray}
where $|k_j \rangle$ denotes the occupation with an $A$ particle
and a $B$ particle in the Schmidt mode $k_j$, and $F(
{k_1,k_2,...,k_N })$ is the weight factor for the state $\left|
{k_1,k_2,...,k_N } \right\rangle$. If these $k$'s have $d$ same
terms (e.g. $k_1=k_2=...=k_d$), and all others terms are distinct,
then the weigh factor should be $N!/d! \times d! =N!$. The $d!$ in
the nominator comes from the fact that $c^{\dag}$ contains
operators $a_k^{\dag} b_k^{\dag}$.  This argument applies to
general sequence with any combinations of degenerate terms.

\end{references}
\end{document}